\documentclass[a4paper]{article}

\usepackage{booktabs} 
\usepackage{multirow} 
\usepackage{threeparttable} 
\usepackage{color}  
\usepackage{amsmath}
\usepackage{caption} 
\usepackage{etoolbox,siunitx}
\usepackage{setspace}

\renewenvironment{thebibliography}[1]{
  \begin{oldthebibliography}{#1}
}
{
  \end{oldthebibliography}
}

\usepackage[
backend=biber,
style=ieee,
citestyle=numeric-comp,
maxbibnames=3,
maxcitenames=3,
doi=false,isbn=false,url=false,eprint=false
]{biblatex}

\defbibheading{bibliography}[\refname]{}

\DeclareSourcemap{
	\maps[datatype=bibtex, overwrite=true]{
		\map{
			\step[fieldsource=publisher,
			match=\regexp{.+},
			replace={{}}]
			\step[fieldsource=month,
			match=\regexp{.+},
			replace={{}}]
			\step[fieldsource=location,
			match=\regexp{.+},
			replace={{}}]
			\step[fieldsource=address,
			match=\regexp{.+},
			replace={{}}]
			\step[fieldsource=organization,
			match=\regexp{.+},
			replace={{}}]
		}
	}
}

\usepackage{INTERSPEECH2022}

\title{ESPnet-SE++: Speech Enhancement for Robust Speech Recognition, Translation, and Understanding}

\name{Yen-Ju Lu$^{*1}$, Xuankai Chang$^{*2}$, Chenda Li$^3$, Wangyou Zhang$^3$, Samuele Cornell$^4$,  Zhaoheng Ni$^5$, Yoshiki Masuyama$^{2,7}$, Brian Yan$^2$, Robin Scheibler$^6$, Zhong-Qiu Wang$^2$, Yu Tsao$^1$, Yanmin Qian$^3$, Shinji Watanabe$^2$ \thanks{$^*$ denotes equal contribution.}}
\address{
$^1$Academia Sinica, Taipei\,\,
$^2$Carnegie Mellon University, USA \\
$^3$Shanghai Jiao Tong University, Shanghai\,\,
$^4$Università Politecnica delle Marche, Italy\\
$^5$Meta AI, USA \,\,
$^6$LINE Corporation, Japan \,\,
$^7$Tokyo Metropolitan University, Japan
}
\email{}

\begin{document}

\ninept
\maketitle
%
\begin{abstract}
  This paper presents recent progress on integrating speech separation and enhancement (SSE) into the ESPnet toolkit.
  Compared with the previous ESPnet-SE work, numerous features have been added, including recent state-of-the-art speech enhancement models with their respective training and evaluation recipes.
  Importantly, a new interface has been designed to flexibly combine speech enhancement front-ends with other tasks, including automatic speech recognition (ASR), speech translation (ST), and spoken language understanding (SLU).
  To showcase such integration, we performed experiments on carefully designed synthetic datasets for noisy-reverberant multi-channel ST and SLU tasks, which can be used as benchmark corpora for future research.
  In addition to these new tasks, we also use CHiME-4 and WSJ0-2Mix to benchmark multi- and single-channel SE approaches. Results show that the integration of SE front-ends with back-end tasks is a promising research direction even for tasks besides ASR, especially in the multi-channel scenario. 
  The code is available online at \url{https://github.com/ESPnet/ESPnet}. The multi-channel ST and SLU datasets, which are another contribution of this work, are released on HuggingFace. 
\end{abstract}

\section{Introduction}
Speech separation and enhancement (SSE) aims at extracting a target speech signal from noise, reverberation, and interfering speakers. It is essential to robust speech recognition~\cite{Rank_1-Wang2018,End_to_end-Zhang2021}, assistive hearing~\cite{wang2017deep}, and robust speaker
recognition~\cite{H.L.Hansen2015}. SSE methods have benefited greatly from the recent development of deep learning (DL) approaches, and DL-based methods are now the de-facto standard. 
This opens up the possibility for end-to-end integration of SSE methods with many downstream speech processing back-end tasks [e.g., automatic speech recognition (ASR), keyword spotting, and speech translation (ST), to name a few].
In fact, many works have been recently exploring this research direction, mainly concerning ASR \cite{Speech-Subramanian2019,MIMO-Chang2019,Location-Subramanian2020,chen2020continuous}. 

To accelerate research in SSE, ESPnet-SE toolkit~\cite{ESPnet_SE-Li2021} was developed and currently supports multiple state-of-the-art enhancement approaches and various corpora. 
In the meantime, there is currently a lot of effort and interest for robust ASR, ST, and spoken language understanding (SLU) in noisy and possibly distant speech scenarios \cite{haeb2019speech,CHiME-4-Vincent2017}, such as ones encountered by smart-speaker devices. 
This motivated us to extend ESPnet-SE into ESPnet-SE++, which has a re-designed interface focused on modularity. This new interface allows for a seamless combination of different front-end SSE models with various downstream tasks such as ASR, ST and SLU. 


The contributions of ESPnet-SE++ are summarized below:
\begin{itemize}
    \item  We significantly extend ESPnet-SE by providing new recipes for various enhancement corpora and challenges. 
    \item  In addition to new recipes, several state-of-the-art single-channel and multi-channel enhancement approaches have been added. Including unsupervised separation~\cite{Unsupervised-Wisdom2020} and generative speech enhancement~\cite{lu2022conditional} approaches. 
    \item A redesigned modular interface allows a flexible, ``plug-and-play" combination of SE front-ends with different ESPnet back-end tasks such as ASR, SLU, ST, etc. As we showcase in Section \ref{sec:combination} for ST and SLU, this easily allows to test and possibly fine-tune (as done for SLU) the same pre-trained front-end for multiple applications.
    \item We develop two multi-channel noisy-reverberant datasets derived from SLURP and Libri-Trans, respectively. We simulate potential applications of SLU and ST in a distant speech setting and showcase how tight integration of SE front-end with the back-end opens up promising research directions for other tasks besides ASR. 
\end{itemize}

An extensive experimental evaluation is performed to showcase the flexibility of ESPnet-SE++. 
In particular, we conducted different experiments using four different datasets: CHiME-4, WSJ0-2mix, and the two purposedely developed SLU and ST multi-channel datasets mentioned previously.
Results show that speech enhancement can improve the performance in a wide variety of downstream tasks. 

\section{Related Works}
In this section, we briefly compare ESPnet-SE with other open-source deep learning-based speech enhancement and separation toolkits, such as \textit{nussl} (North-western University Source Separation Library) \cite{nussl}, \textit{Onssen} (An Open-source Speech Separation and Enhancement Library)~\cite{ni2019onssen}, \textit{Asteroid} (Audio source separation on Steroids)~\cite{Pariente2020Asteroid}, and \textit{SpeechBrain}~\cite{speechbrain}. While \textit{nussl} provides several state-of-the-art SE methods, the data preparation and experiments of \textit{nussl} and \textit{Onssen} are not easily configurable from the command line~\cite{Pariente2020Asteroid}. On the other hand, \textit{Asteroid} and \textit{SpeechBrain} provide a whole pipeline from data preparation to enhancing and evaluating the testing speech.

\begin{figure}[t]
  \begin{minipage}{\linewidth}
    \centering
    \centerline{\includegraphics[width=\textwidth]{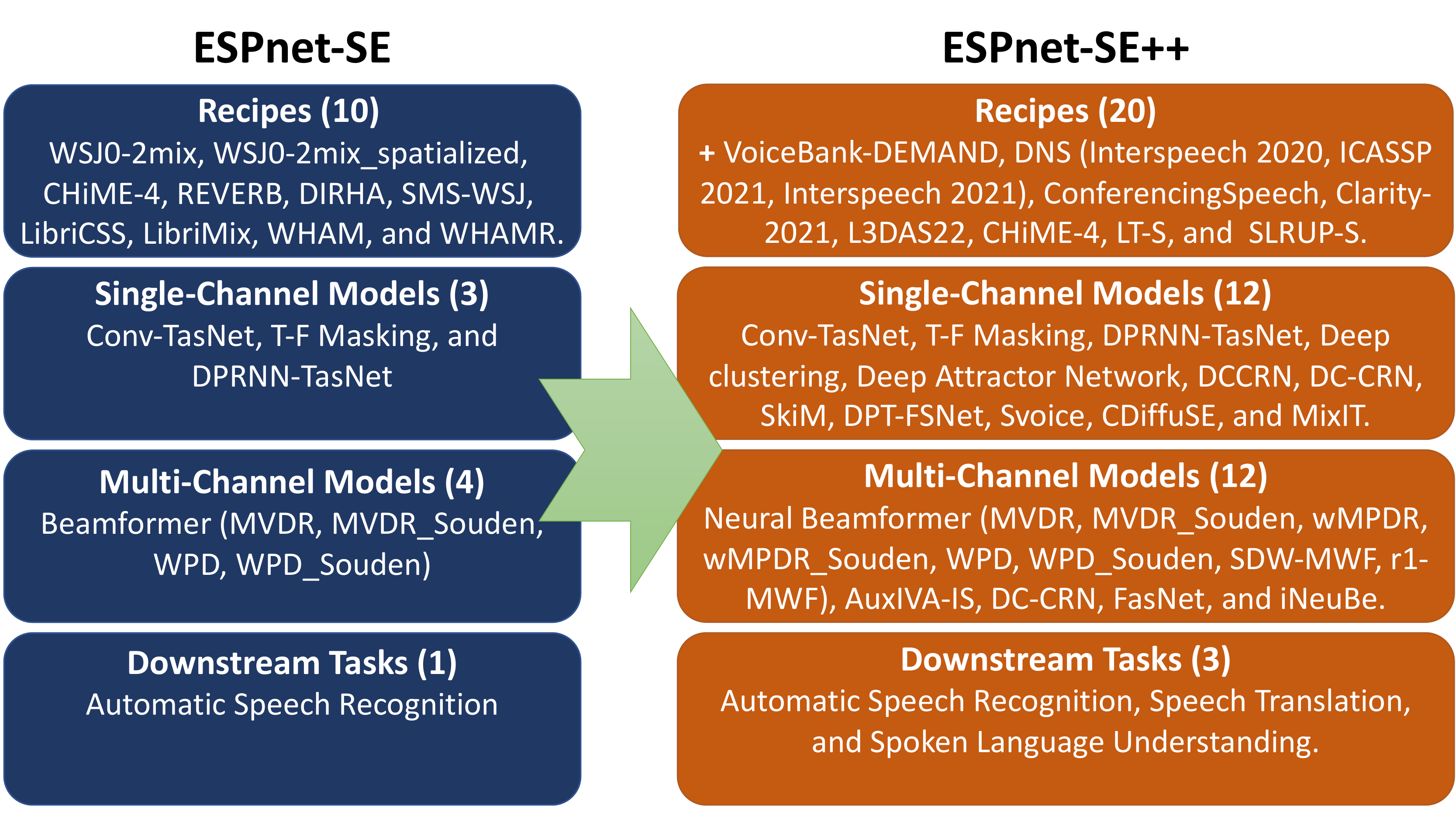}}
  \end{minipage}
\caption{Comparison for ESPnet-SE and ESPnet-SE++. Recipes in ESPnet-SE are not shown in the block of ESPnet-SE++.}
\label{fig:SE++ compare}
\end{figure}

However, \textit{Asteroid} is exclusively designed for SE front-end processing. On the other hand, \textit{SpeechBrain}, supports both SSE front-end and back-end (e.g. ASR) techniques along with corresponding recipes but does not yet integrate multiple speech processing tasks into a single recipe. Although the application of SE can already be beneficial, especially in the multi-channel case, various studies suggest that joint-optimization with downstream tasks can further boost the performance~\cite{haeb2019speech,von2020multi,Closing-Zhang2021}.
We summarize the progress between ESPnet-SE and ESPnet-SE++ in Figure~\ref{fig:SE++ compare}. ESPnet-SE++ owns 20 SSE recipes, with 24 different enhancement/separation models.

\section{Features of ESPnet-SE++}

ESPnet-SE covers a wide range of speech enhancement/separation models and recipes, including (1) single-channel and multi-channel; (2) single-source and multi-source; (3) time-domain and frequency-domain enhancement models. ESPnet-SE++, is extended to support the latest state-of-the-art models and the latest enhancement/separation recipes, and also provides the combination of enhancement with other downstream tasks, including ASR, ST, and SLU. In addition, ESPnet-SE++ supports the native complex datatype built by PyTorch for complex time-frequency domain processing instead of relying on custom workarounds built on top of PyTorch real-valued tensors. 



\subsection{Recipes}
\label{Sec.recipes}
\begin{figure}[t]
  \begin{minipage}{\linewidth}
    \centering
    \centerline{\includegraphics[width=\textwidth]{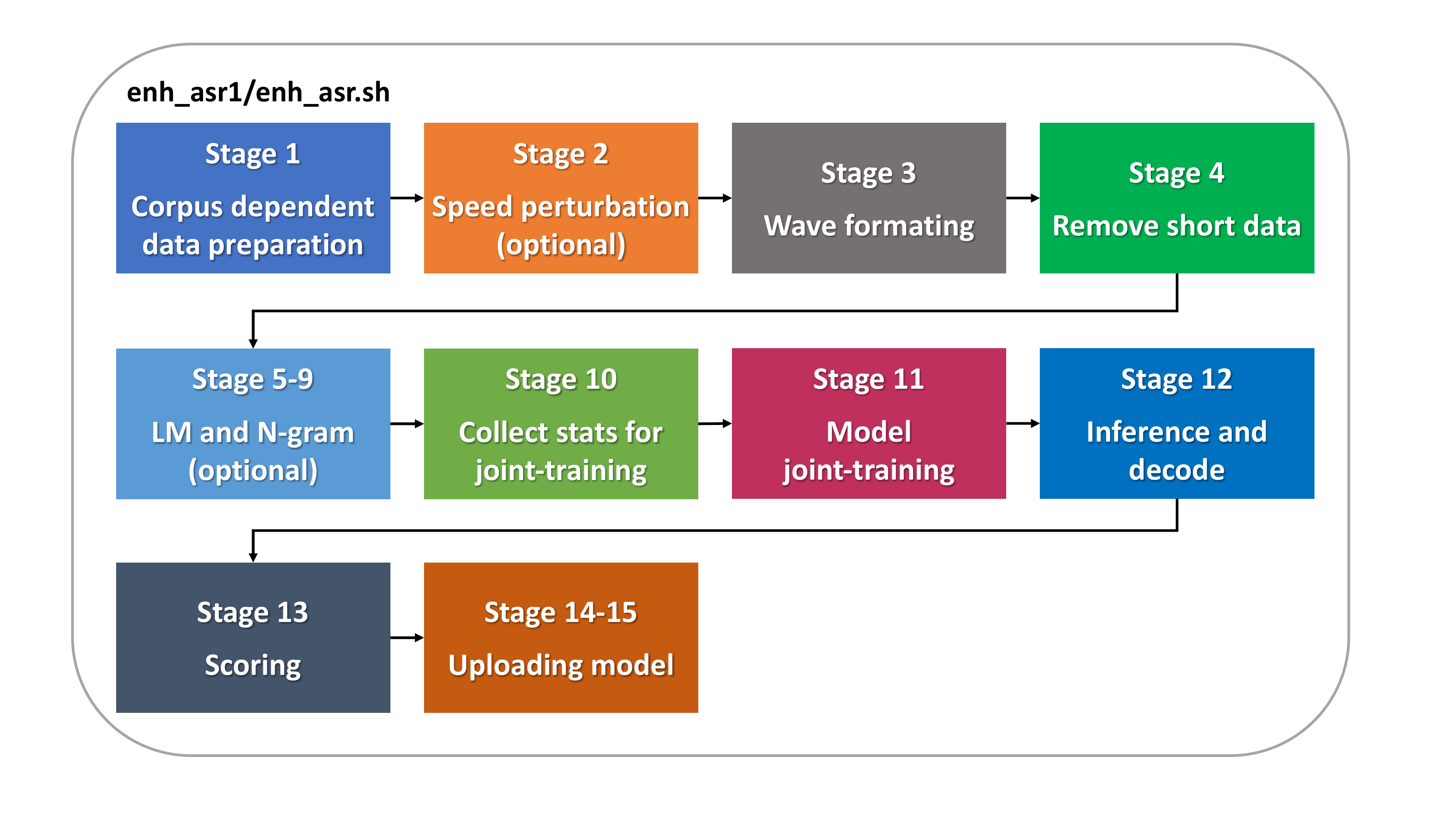}}
  \end{minipage}
\caption{Block diagram of \texttt{enh$\_$asr.sh} in ESPnet-SE++, the combination of SE and back-end models.}
\label{fig:SE++ diagram}
\end{figure}

ESPnet-SE provides various recipes for several SSE benchmark corpora which are derived from WSJ \cite{LDC-LDC1993} and Librispeech \cite{Librispeech-Panayotov2015} corpora.
ESPnet-SE++, on top of these, adds $10$ new recipes, including speech enhancement/separation corpora: Voicebank-DEMAND~\cite{Investigating-Valentini-Botinhao2016}, speech enhancement challenges: DNS (Interspeech 2020~\cite{DNS2020IS-Reddy20}, ICASSP 2021~\cite{DNS2021ICASSP-Reddy2021}, and Interspeech 2021~\cite{DNS2021IS-Reddy21}), ConferencingSpeech 2021~\cite{Conferencingspeech-Rao2021}, Clarity-2021~\cite{Clarity-Graetzer2021}, and L3DAS22~\cite{L3DAS22-Guizzo2022}. 
In addition, we develop a new recipe that enables an all-in-one combination of multiple speech processing components with a front-end SSE module. This includes robust ASR (CHiME-4), ST (Libri-Trans~\cite{LibriTrans-kocabiyikoglu2018} mixture), and SLU (SLURP~\cite{SLURP-Bastianelli2020} mixture) datasets. These recipes contain the data preparation steps for both enhancement and downstream speech processing tasks, the joint-training of the front-end and back-end models, and the output of enhanced/separated speech with the recognition/translation/understanding results.
The function of each stage in the unified pipeline is depicted in Figure~\ref{fig:SE++ diagram}.

\subsection{Models}
In ESPnet-SE++, we add various classical and state-of-the-art supervised SSE models, including (1) single-channel models: Deep Clustering~\cite{Deep-Hershey2016}, Deep Attractor Network~\cite{Deep-Chen2017}, DCCRN~\cite{DCCRN-Hu2020}, DC-CRN~\cite{Deep-Tan2021}, SkiM~\cite{li2022skim}, DPT-FSNet~\cite{dang2021dpt}, SVoice~\cite{SVoice-Nachmani2020}, and (2) multi-channel models: neural beamformer with more supported types (c.f. Section~\ref{ssec:multich}), FaSNet~\cite{FaSNet-Luo2019}, and iNeuBe~\cite{wang2021leveraging,iNeuBe-Lu2022}.
In addition, we further extend the ESPnet-SE framework to support various training and inference procedures, including (3) generative enhancement model, CDiffuSE~\cite{lu2022conditional}, and (4) unsupervised separation model, MixIT~\cite{Unsupervised-Wisdom2020}.

 

\subsection{Training Objectives}
\label{ssec:loss}
The training objective can vary a lot between different speech enhancement and separation tasks.
To make it more flexible and customizable, we disentangle the training objective into two abstract classes: a \textit{criterion} and a \textit{wrapper}. 

The \textit{criterion} is an implementation of a simple loss function, e.g. SI-SNR \cite{luoTaSNetTimeDomainAudio2018,SDR__Half_baked-LeRoux2019}, CI-SDR \cite{Convolutive-Boeddeker2021,Scheibler_fast_bss_eval}, mean square error (MSE) on time-frequency masks, etc.
These losses usually take both target and estimated tensors as input and output a scalar loss value.

The \textit{wrapper} instead deals with a post-processing of the \textit{criterion} output. For example, the procedure to find the best permutation in the PIT \cite{Multitalker-Kolbaek2017} algorithm is implemented as a \textit{wrapper}.
We have also added MixIT \cite{Unsupervised-Wisdom2020} \textit{wrapper} for unsupervised speech separation.
This modular design allows researchers to add complex training objectives conveniently by writing custom \textit{wrapper} classes and combining it with a suitable \textit{criterion}.

Furthermore, ESPnet-SE++ supports multiple training objectives. Different training objectives can be combined with different weights in a multi-task learning (MTL) fashion. The specific implementation of training objectives, including their \textit{wrapper}, \textit{criterion} and MTL weight, can be selected and configured directly in the ESPnet \textit{yaml} configuration file.


\subsection{Single- and Multi-channel Speech Enhancement}
\label{ssec:multich}
As mentioned above, ESPnet-SE++ adds several models on top of ESPnet-SE. 
For single-channel models, DCCRN~\cite{DCCRN-Hu2020} and DC-CRN~\cite{Deep-Tan2021} are added.
To catch up with the latest development of deep learning-based multi-channel speech enhancement systems, we further added several multi-channel SSE models, including FaSNet~\cite{FaSNet-Luo2019}, DC-CRN~\cite{Deep-Tan2021}, and iNeuBe~\cite{wang2021leveraging,iNeuBe-Lu2022}.
The first one is a time-domain multi-channel model, while the other two operate in the complex time-frequency domain.
In addition, there are also various notable updates to the existing ESPnet-SE neural beamformer model.
Firstly, the \texttt{DNN\_Beamformer} module is refactored to be more user-friendly, while keeping backward compatibility.
Secondly, more beamformer variants are supported, such as weighted minimum power distortionless response (wMPDR)~\cite{Jointly-Boeddeker2020}, speech distortion weighted multi-channel Wiener filtering (SDW-MWF)~\cite{Spatially-Spriet2004}, and rank-1 multi-channel Wiener filter (r1-MWF)~\cite{Rank_1-Wang2018} etc.
A differentiable implementation of the blind source separation algorithm AuxIVA-ISS~\cite{scheiblerSurrogateSourceModel2021} has also been added.
Thirdly, these beamforming solutions now rely on the native complex tensors data-type built in PyTorch 1.9.0+, which shows a faster speed and comparable performance compared to our previously-adopted implementation.~\footnote{\url{https://github.com/kamo-naoyuki/pytorch_complex}}
We also integrate multi-channel processing methods provided by TorchAudio~\cite{yang2021torchaudio} such as power spectral density (PSD) computation, MVDR beamforming, etc.~\footnote{\url{https://pytorch.org/audio/main/functional#multi-channel}}
Last, several numerical stability-related techniques proposed in~\cite{End_to_end-Zhang2021} have been integrated to improve the training stability and performance.

In Table~\ref{tab:exp_CHiME-4}, we list the SSE performance of various models supported in the current toolkit on the CHiME-4 corpus, where the newly-added models are marked with $^\star$.

\begin{table}[t]
\centering
\caption{Results of single- and multi-channel speech enhancement approaches on CHiME-4.}
\label{tab:exp_CHiME-4}
\begin{threeparttable}
\resizebox{\columnwidth}{!}{
\begin{tabular}{ll|ccc}
\toprule
Models & & PESQ & STOI & SI-SNR (dB) \\ \hline
\multicolumn{2}{l|}{No processing} & 2.18 & 0.870 & 7.51 \\ 
\multicolumn{2}{l|}{Wang \emph{et al.}~\cite{wang2020complex}} & \textbf{3.68} & \textbf{0.986} & \textbf{22.00} \\ \hline
\multicolumn{5}{c}{\emph{Single-channel Models}} \\ \hline
\multicolumn{2}{l|}{Conv-TasNet (baseline)} & 2.58 & 0.892 & 11.57 \\
\multicolumn{2}{l|}{$^\star$DCCRN} & \textbf{2.59} & \textbf{0.895} & \textbf{12.57} \\
\multicolumn{2}{l|}{$^\star$DC-CRN} & 2.43 & 0.880 & 11.59 \\ \hline
\multicolumn{5}{c}{\emph{Multi-channel Models}} \\ \hline
\multirow{8}{*}{\shortstack[l]{Neural\\Beamformer}} & MVDR & 2.61 & \textbf{0.954} & 13.98 \\
& MVDR\_Souden & 2.66 & \textbf{0.954} & \textbf{15.25} \\
& $^\star$wMPDR & 2.60 & 0.951 & 13.53 \\
& $^\star$wMPDR\_Souden & 2.64 & 0.951 & 15.24 \\
& WPD & 2.60 & 0.950 & 13.43 \\
& WPD\_Souden & 2.64 & 0.950 & 14.89 \\
& $^\star$SDW-MWF & 2.43 & 0.922 & 11.87 \\
& $^\star$r1-MWF & \textbf{2.67} & 0.953 & 15.08 \\ \hline
 \multicolumn{2}{l|}{$^\star$AuxIVA-ISS} & 2.49 & 0.900 & 10.34 \\\hline
\multicolumn{2}{l|}{$^\star$DC-CRN} & 2.95 & 0.948 & 17.04 \\
\multicolumn{2}{l|}{$^\star$FaSNet} & 2.70 & 0.935 & 14.83  \\
\multicolumn{2}{l|}{$^\star$iNeuBe (DNN$_1$)} & \textbf{3.24} & \textbf{0.969} & \textbf{19.52} \\ \bottomrule
\end{tabular}
}
\begin{tablenotes}[flushleft]\footnotesize
    \item[*] The symbol $^\star$ denotes models newly-added to ESPnet-SE++.
\end{tablenotes}
\end{threeparttable}
\end{table}


\subsection{Single-channel Speech Separation}


In the initial version of ESPnet-SE \cite{ESPnet_SE-Li2021}, we implemented some time-frequency (TF) domain \cite{Multitalker-Kolbaek2017} and time domain models \cite{Conv_TasNet-Luo2019,Dual_path-Luo2020}.
In the new implementation, we design these two kinds of models into a unified framework, which consists of an \textit{encoder}, a \textit{separator} and a \textit{decoder}. 
Depending on the model type, \textit{encoder} and \textit{decoder} could be short-time Fourier transform (STFT) and inverse STFT (iSTFT) for the TF domain models; or they could be convolutional layers and transposed convolutional layers for time domain models such as \cite{luoTaSNetTimeDomainAudio2018}.
The \textit{separator} is typically a sequence mapping neural network. It takes the input from the \textit{encoder}, and generates $S$ output features.
$S$ is the number of speech sources to be separated, and for most speech enhancement models, $S = 1$.
The \textit{decoder} transforms the features into the target audios.

An interface holding together the \textit{encoder}, \textit{separator} and \textit{decoder} has been designed. This modular design allows to explore many different architectural variations with less boilerplate code.
Based on the unified framework, we also enrich the speech separation models in ESPnet-SE, including deep clustering~\cite{Deep-Hershey2016}, deep attractor network~\cite{Deep-Chen2017}, DC-CRN~\cite{Deep-Tan2021}, SkiM~\cite{li2022skim} and, MixIT~\cite{Unsupervised-Wisdom2020}.
The reproduced results on the WSJ0-2mix~\cite{Deep-Hershey2016} benchmark are listed in Table~\ref{tab:exp_wsj0-2mix}.

\begin{table}[t]
\centering
\caption{Results of single-channel speech separation models on WSJ0-2mix.}
\label{tab:exp_wsj0-2mix}
\begin{threeparttable}
\begin{tabular}{l|ccc}
 \toprule
Models & PESQ & STOI & SI-SNR (dB) \\ \hline
No processing & 2.01 & 0.738 & 0.00 \\
Conv-TasNet (baseline) & 3.25 & 0.953 & 15.94 \\ 
DPRNN-TasNet & \textbf{3.47} & \textbf{0.968} & 17.91 \\ 
$^\star$Deep Clustering & 2.15 & 0.845 & 7.91 \\
$^\star$Deep Attractor Network & 2.68 & 0.893 & 10.30 \\
$^\star$DC-CRN & 3.11 & 0.935 & 13.01 \\ 
$^\star$SkiM & \textbf{3.47} & 0.966 & \textbf{18.45} \\  
$^\star$MixIT (Conv-TasNet) & 3.00 & 0.938 & 13.50 \\  \bottomrule
\end{tabular}
\begin{tablenotes}[flushleft]\footnotesize
    \item[*] The symbol $^\star$ denotes models newly-added to ESPnet-SE++.
\end{tablenotes}
\end{threeparttable}
\end{table}






\section{Combination tasks of ESPnet-SE++}\label{sec:combination}
ESPnet-SE++ allows for a tight and easy integration of SSE front-end processing techniques and back-end tasks as introduced in Sec.~\ref{Sec.recipes}. To showcase this, we provide three combined recipes, CHiME-4, LT-S, and SLURP-S,  with speech enhancement (SE) front-end subtask, followed by a back-end ASR/ST/SLU subtask. We performed several experiments with different techniques to assess if and how SE could improve the results of the back-end tasks even when multi-condition training is employed. 



\begin{table}[t]
\centering
\caption{WER for system combination of speech enhancement with speech recognition on CHiME-4 corpus.}
\label{tab:exp_CHiME-4_asr}
\begin{tabular}{l|cc} 
\toprule
Models &  SIMU  (\%) & REAL (\%) \\ \hline
No processing &  19.7  & 18.0 \\ \hline
\multicolumn{3}{c}{\emph{Single-channel Models}} \\ \hline
Conv-TasNet   & 17.4 & \textbf{15.2} \\
DCCRN &  \textbf{16.3}  & 15.8 \\ \hline
\multicolumn{3}{c}{\emph{Multi-channel Models}} \\ \hline
FaSNet & 15.7  & 23.8 \\ 
Neural Beamformer  & 10.8 & \textbf{13.7} \\ 
iNeuBe (DNN$_1$) & \textbf{9.0} & 35.8 \\ 
\bottomrule
\end{tabular}
\end{table}

\subsection{Data Simulation}\label{sec:simulated_data}
We created two multi-channel noisy-reverberant datasets based on SLURP and Libri-Trans (LT): SLURP-S and LT-S , where S stands for spatialized. 
We augmented the original SLURP and LT datasets using room impulse responses generated via Pyroomacoustics \cite{scheibler2018pyroomacoustics}, and noises from FSD50k \cite{fonseca2022fsd50k}, SINS \cite{dekkers2017sins}.
In detail, for each original utterance from SLURP and LT we simulate a smart-speaker scenario where the target signal is captured by a 4-microphone circular array with a diameter of 10\,cm. For each utterance, we sample a room size from uniform distribution $U(10, 100)$ $m^2$ and a reverberation time (T60) from $U(0.2, 0.6)$ s, typical of most indoor settings \cite{traer2016statistics}. Room height is sampled from $U(2.5, 4)$ m.
The target speech is contaminated by $1$ up to $4$ point-source localized noises from FSD50k and, in addition, also one diffuse noise audio sample from SINS. Diffuse noise is simulated using the technique outlined in \cite{habets2008generating}. 
The positions of the array, point-source noises and target are sampled randomly in the virtual room with the constraint of being at least $0.5$\,m apart from each other and the walls. The Signal-to-Noise ratio (SNR) for target versus point-source noises is sampled from $U(0, 15)$ dB, while for diffuse noise is sampled from $U(12, 35)$ dB. 

Regarding SLURP, since it already comprises noisy-reverberant examples, we use DNSMOS model \cite{reddy2021dnsmos} to select only a subset of sufficiently clean utterances (with estimated BAK $\geq 3.2$) from training real, development and test, to use for SLURP-S. 
This process leaves 24497 utterances for \texttt{training\_real}, and 6387 and 6099  for development and test respectively. No utterance is discarded from \texttt{training\_synthetic}. 
The discarded utterances are kept but treated as single-channel examples with no oracle SE supervision, and are left for possible future work on semi/self-supervised joint SE+SLU. 
Instead, LT-S retains the same number of utterances of the original LT as it features only clean speech. 

Finally, in order to assess generalization to unseen noise conditions, we also generate, for both datasets, an additional test set with diffuse noise derived from QUT\,\cite{dean2010qut} instead of SINS.




\subsection{Combination of SE with ASR on CHiME-4}
\label{sec:se-asr}
In Table~\ref{tab:exp_CHiME-4_asr} we report the results obtained by combining different SE front-ends with an ASR model on the CHiME-4 corpus~\cite{CHiME-4-Vincent2017}.
CHiME-4 consists of both real and simulated noisy speech signals and is particularly suited for this analysis.
We used a transformer encoder-decoder as the E2E ASR model that is pre-trained using the official ESPnet CHiME-4 recipe.
The SE models are instead pre-trained using simulated data. The performance of these systems in terms of signal-based metrics is summarized in Sec.~\ref{ssec:multich}.
After pre-training, the entire SE+ASR system is fine-tuned together with both SE and ASR losses.
On the simulated data, all SE+ASR systems improve performance with respect to using solely the ASR model.
This result confirms the effectiveness of jointly fine-tuning SE and ASR systems for robust speech recognition.
Meanwhile, some models resulted in worse WERs on the real data.
This is a commonly observed phenomenon when evaluating SE models on simulated and real data~\cite{Human-Eskimez2021,Closing-Zhang2021}, especially when they are not jointly optimized with the downstream ASR task.
The neural beamformer achieved the best performance in both simulated and real sets.

\begin{table}[tb]
    \centering
    \sisetup{table-format=2.1,round-mode=places,round-precision=1,table-number-alignment = center,detect-weight=true}
    \caption{Results of combining speech enhancement with speech translation on LT-S corpus and speech enhancement with spoken language understanding on SLURP-S corpus.}
    \resizebox{\columnwidth}{!}{
    \begin{tabular}{l|cc|cc}
        \toprule
        \multirow{2}{*}{SE} & \multicolumn{2}{c|}{LT-S (BLEU)} & \multicolumn{2}{c}{SLURP-S (Acc\%)} \\
        \cmidrule{2-5}
         & TEST & TEST QUT & TEST & TEST QUT \\
        \midrule
        No processing & 7.3 & 7.4 & 74.2 & 73.4 \\
        \midrule
        \multicolumn{5}{c}{Single-channel Models} \\
        \midrule
        Conv-TasNet & 7.4 & 7.5 & 64.3 & 64.0 \\
        DCCRN & \textbf{10.1} & \textbf{10.1} & \textbf{77.0} & \textbf{76.3} \\
        \midrule
        \multicolumn{5}{c}{Multi-channel Models} \\
        \midrule
        FaSNet & 7.7 & 3.6 & 73.1 & 72.5  \\
        iNeuBe (DNN$_1$) & 13.2 & 13.2 & 77.8 & 77.5 \\
        iNeuBe (mfmcwf) & 13.0 & 13.0 & 80.1 & 80.2 \\
        iNeuBe (DNN$_2$) & \textbf{14.4} & \textbf{14.6} & \textbf{80.4} & \textbf{80.3} \\
        \bottomrule
    \end{tabular}
    \label{tab:exp_se_st_slu}
    }
\end{table}

\vspace{-5pt}
\subsection{Combination of SE with ST on LT-S}
\label{ssec:se-st}


In the left part of Table~\ref{tab:exp_se_st_slu}, we explore the combination of SE with a ST task performed on the simulated LT-S dataset outlined in Section \ref{sec:simulated_data}. For the ST model, we used a conformer-encoder transformer-decoder as the back-end model. The ST model was trained on the original clean Libri-Trans dataset and achieved a BLEU score of 16.7 on the test set. The front-end SE models were pre-trained using both LT-S and SLURP-S datasets using the anechoic clean speech as target (thus performing joint denoising and dereverberation). For the evaluation, we directly concatenate the pretrained SE models with the ST model in an end-to-end manner. Note that the whole system was not fine-tuned.
We can see that the addition of the SE front-end generally improves the performance against the noisy speech even without fine-tuning. Among the SE models considered, both iNeuBe models based on DNN$_2$ and multi-frame multi-channel Wiener filter (mfmcwf), bring the largest performance gain. In the single-channel case, DCCRN, as expected, outperforms Conv-Tasnet.

\vspace{-6pt}
\subsection{Combination of SE with SLU on SLURP-S}
\label{ssec:se-slu}
In the right part of Table~\ref{tab:exp_se_st_slu}, we explore instead the combination of SE with a SLU back-end and report the intent classification accuracy.
Regarding the SLU model, we used the conformer-encoder transformer-decoder trained with multi-condition training (clean and noisy-reverberant) on the SLURP-S corpus.
After SLU back-end pre-training, we fine-tuned the SE and SLU systems together. The SE models are the same as used in the ST experiments in Sec.~\ref{ssec:se-st}.
Notably, we can see that in this instance, contrary to what is observed on CHiME-4 which is arguably less acoustically challenging than SLURP-S, Conv-TasNet and FaSNet lead to degraded performance. On the other hand, both DCCRN and iNeuBe are able to outperform the multi-condition SLU model. This confirms that models that rely on complex spectral mapping \cite{wang2020complex, iNeuBe-Lu2022} are more robust in challenging acoustical conditions. 

\vspace{-6pt}
\section{Conclusions}
In this work we presented the ESPnet-SE++ toolkit.
Built on top of ESPnet-SE, it includes new state-of-the-art  models, losses, and recipes for speech enhancement corpora and challenges in various scenario. 
The toolkit interface has also been overhauled and improved with greater modularity. This allows a flexible combination of speech enhancement with ESPnet back-end tasks such as ASR, ST, and SLU. 
As an additional contribution, to showcase such integration of front-end and back-end components, we developed two multi-channel noisy-reverberant corpora based on SLURP and Libri-Trans.
Experiments show that the use of SE front-end models improves both signal-based evaluation metrics and back-end tasks such as ASR, ST, and SLU, even when the back-end is trained on a multi-condition fashion. 
Future work could investigate further integration of different SE techniques with back-end tasks, including self-supervised pre-training via MixIT and generative enhancement models, with the goal of improving generalization to unseen acoustical conditions. 



\section{References}
{
\begingroup
\setstretch{0.8}
\setlength\bibitemsep{0.95pt}
\printbibliography
\endgroup
}

\end{document}